# Connected Big Data Measurement

Rossi Kamal, *Student Member, IEEE,* Choong Seon Hong, *Senior Member, IEEE,*

*Abstract*—In this paper, we have summarized how resilient Big Data monetization scheme outperforms state-of-the art schemes by maintaining a balance between CDS size and routing.

## I. Experiments

### A. Environment

The proposed scheme is evaluated through an experimental Java implementation. Hence, nodes are randomly generated by keeping x and y-coordinates 20-500. In this process, 10 UDGs are randomly generated by keeping randomly generated nodes as inputs of these connected graph. Finally, CDS is constructed for each of the UDG by applying these schemes. It is notable that unit disk graph comprises of nodes having equal transmission range and assuming cost any obstacle. In this context, n is varied from 1 to 10 by varying transmission range from 10 to 40. In this process, 20 randomly generated graph instances are averaged out to conclude any decision.

### B. Methodology

*Hop-dist (a, b)*: The shortest path between a pair of nodes $a$ to $b$ is regarded as hop-distance and denoted as $d(a, b)$ and $d_D(a, b)$ for original graph $G$ and dominating set $D$, respectively.

*Diameter (a,b): [1]* The diamter is the maximum Hop-dist $d_{D_{max}}(a,b)$ in a dominating set, where $(a, b) \epsilon D$.

*ABPL (a,b): [2]* The ABPL of a dominating set is the average Hop-dist $d_{D_{avg}}(a, b)$, where $a, b \epsilon D$.

*α-MOC-path(a, b): [3]* α-MOC-path represents path in a dominating set, such that the number of intermediate nodes is α-times than on the shortest path in the original graph.

*Collaborative-cover (a, b)[4]* : Collaborative cover represents highest effective cover $\frac{a}{b}$, which represents cover on $b$ by $a$ in its ($b$) one-hop vicnity.

*Guaranteed path (a, b): [5]* Guaranteed path represents path between pair of nodes $a, b$, such that $d_D(a, b) \leq 7d(a, b)$

*Resilient path (a, b): [6][7]* Resilient path represents path between pair of nodes $a$ and $b$, such that $d_D(a, b) \leq 5d(a, b)$.

### C. Strategy

The larger the CDS size is, the more is the chance of involvement of nodes. On the contrary, the smaller the CDS is, the more is the chance of missing routing properties. Eventually, routing through CDS consequently becomes quite larger than shortest hops in original graphs. Thus, the absence of shortest path is a disadvantage even though many benefits are achieved with CDS size. Hence, CDS size is needed to be as small as possible by keeping routing length in minimal stage. Consequently, experiment is intended to acquire desired trade-off between CDS size and routing length.

### D. Results

*1) CDS Size:* Fig.1 shows that proposed scheme outperforms guaranteed, alpha-moc, abpl and diameter-based schemes in terms of CDS size. However, as network becomes dense, gradual performance improvement is observed with guaranteed, alpha-moc and abpl-based schemes for the availability of more intermediate nodes in between dominators. On the contrary, proposed scheme seems to gradually compromise CDS size, which is a trade-off of its relatively strict routing constraint. However, diameter based scheme gradually shows worse performance due to intense focus on largest path.

As number of nodes increases, CDS size increases. Because, the availability of nodes necessitates the addition of more nodes to CDS. However, as network becomes dense, gradual performance improvement is achieved in terms of CDS size. Because, this increases the availability of intermediate nodes in between dominators. Maintaining a well-balance between CDS size, routing and lifetime is a major goal of the proposed scheme. Therefore, slightly increased CDS size is acceptable, as long as routing path and lifetime are maintained within our desired bound.

Fig.2 shows that resilient scheme outperforms guaranteed scheme and other schemes having bit and significant differences respectively. As vertex cardinality increases, more domination-reluctant nodes appear in between dominators. Therefore, CDS size gradually decreases with the increase in vertex cardinality.

*2) Routing Length:* Maximum routing path length (MRPL) represents the worst case routing path. Hence, the shorter the MRPL is, the better is the efficacy of any routing scheme. Fig. shows that the proposed scheme outperforms conventional schemes in terms of MRPL. Despite the fact that, our scheme aims at facilitating a well-balance between CDS size, routing and lifetime, significant performance supremacy is observed with MRPL over the contemporary schemes. .On the other hand, Average Routing Path Length (ARPL) measures aver- age performance of any routing scheme. Fig shows that the proposed scheme ends up with reduced average routing path in comparison to other schemes. As number of node grows or transmission range increases, the performance difference between the proposed scheme and other schemes becomes significant.

In summary, as number of total nodes are small in a connected network, routing path increases at a higher rate

R. Kamal,C.S.Hong is with the Department of Computer Engineering, Kyung Hee University, South Korea e-mail: (see http://www.michaelshell.org/contact.html).

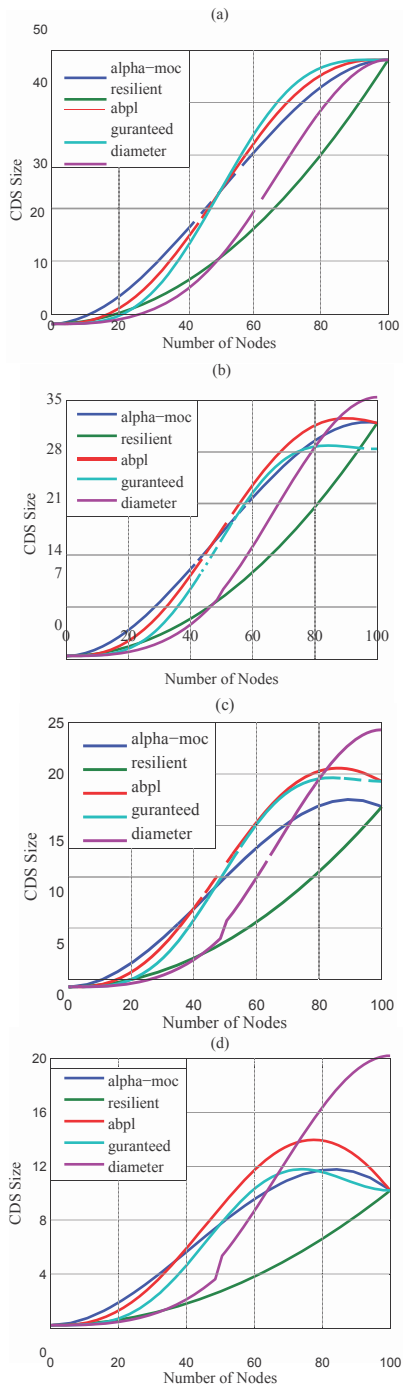

Fig. 1. CDS size comparison by varying number of nodes for different transmission ranges (a) 10, (b) 20, (c) 30, and (d) 40

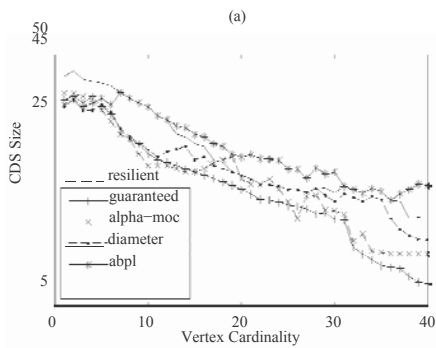

Fig. 2. CDS size observation by varying vertex cardinality

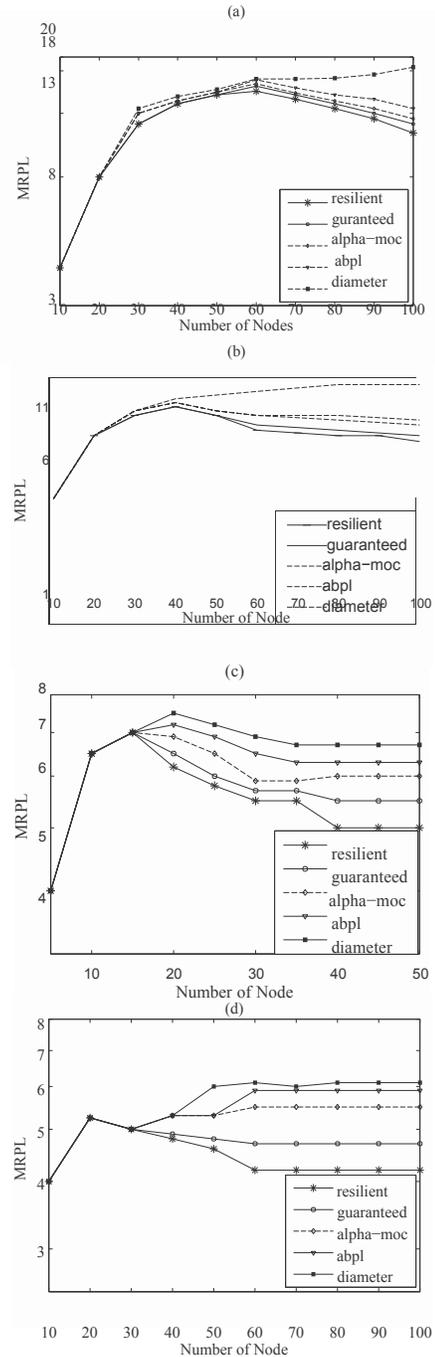

Fig. 3. MRPL(Maximum Routing Path Length) comparison by varying number of nodes for different transmission ranges (a) 10, (b) 20, (c) 30, and (d) 40

with the inclusion of nodes. However, total number of nodes increase to a higher value, they seem to be connected. As a result, distance between them gradually becomes small. Therefore, routing path length gradually decreases, as network size becomes large. As transmission range increases, networks become more connected. Therefore, routing path is expected to decrease, as transmission range increases.

Finally, the proposed scheme seems to supreme over conventional techniques in terms of routing path, with a slightly tradeoff for larger CDS, especially at large-scales. Hence, this matches the justification of supremacy of the proposed scheme over contemporary one claimed in the theoretical analysis

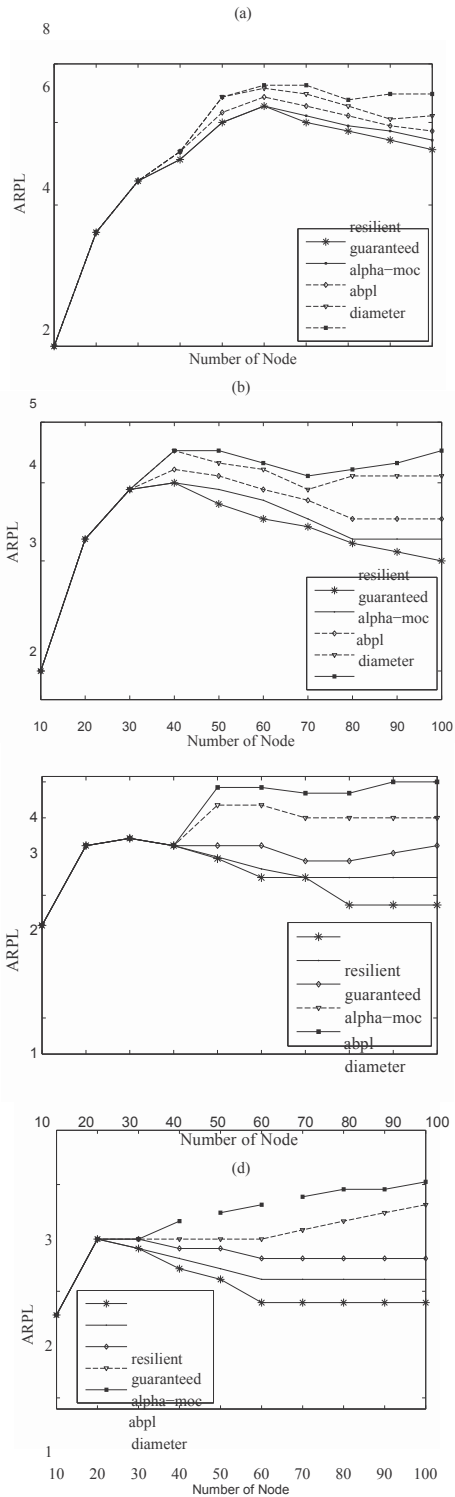

Fig. 4. ARPL(Average Routing Path Length) comparison by varying number of nodes for different transmission ranges (a) 10, (b) 20, (c) 30, and (d) 40

section.

# APPENDIX A
## PROOF OF THE FIRST ZONKLAR EQUATION

Appendix one text goes here.

# APPENDIX B

Appendix two text goes here.

# ACKNOWLEDGMENT

The authors would like to thank...